\documentclass[aps,pre,preprint,showpacs,nofootinbib,superscriptaddress]{revtex4-1}
\usepackage{graphicx}
\usepackage{amsmath}
\usepackage{bm}
\usepackage{amsfonts}
\usepackage{amssymb}

\renewcommand{\vec}[1]{\mathbf{#1}}
\newcommand{\unitvec}[1]{\mathbf{\hat{#1}}}
\newcommand{\matr}[1]{\hat{#1}}

\newcommand{\transpose}[1]{\mathop{{#1}^t}\nolimits}
\newcommand{\Id}{\matr{\mathbb{I}}}
\newcommand{\matE}{\matr{\mathcal{E}}}

\newcommand{\rmi}{\mathrm{i}}
\newcommand{\rme}{\mathrm{e}}
\newcommand{\rmo}{\mathrm{o}}

\begin{document}

\title{Entanglement and chaos in the kicked top}
\author{M. Lombardi}
\affiliation{Laboratoire de Spectrom\'{e}trie Physique (CNRS Unit\'{e} 5588), Universit%
\'{e} Joseph-Fourier Grenoble-1, BP 87, 38402 Saint-Martin d'H\`eres, France}
\author{A. Matzkin}
\affiliation{Laboratoire de Physique Th\'{e}orique et Mod\'{e}lisation (CNRS Unit\'{e}
8089) \\
Universit\'{e} de Cergy-Pontoise, Site de Saint Martin, 95302 Cergy-Pontoise
cedex, France}

\begin{abstract}
The standard kicked top involves a periodically kicked angular momentum.
By considering this angular momentum as a collection of entangled spins, we compute the
bipartite entanglement dynamics as a function of the dynamics of the
classical counterpart. Our numerical results indicate that the entanglement
of the quantum top depends on the specific details of the dynamics
of the classical top rather than depending universally on the global properties of the classical regime.
These results are grounded on linking the entanglement rate to averages
involving the classical angular momentum, thereby explaining why regular dynamics can
entangle as efficiently as the classically chaotic regime. The findings are in line with previous results obtained with a 2-particle top model, and we show here that the standard kicked top can be obtained as a limiting case of the 2-particle top.

\end{abstract}

\maketitle

\section{Introduction}

The quantum-classical correspondence is the hallmark of semiclassical
systems.\ These are genuine quantum systems for which the semiclassical
expansion in the path integral propagator holds (at least for some
appropriately chosen dynamical or temporal regimes). The quantum-classical
correspondence allows to compute and interpret the properties of a quantum
system in terms of the properties of its classical counterpart \cite{gutz}.
This is particularly important for systems displaying a complex dynamics,
for which exact quantum computations are either unfeasible or yield
numerical results that hardly give any clues allowing to grasp the dynamics
of the quantum system.

These last few years several studies aiming to apply the quantum-classical
correspondence to the understanding of dynamical entanglement have been
published \cite{nemes99,fujisaki03,fujisaki04,epl06,laksha,zhang08,chung09}.
Although entanglement is a distinctive quantum feature without
a classical counterpart, many quantum systems displaying entanglement have a
classical counterpart. The idea in these type of studies is then to assess
whether there is a link between the generation of entanglement in the
quantum system and the underlying classical dynamics. It was initially
suggested that underlying chaotic dynamics was correlated with higher and
faster entanglement. However it was later realized that integrable dynamics
could lead to equivalent or even more efficient entanglement \

In earlier works \cite{epl06,pra06}, we have shown that the entanglement
dynamics did depend on classical phase-space features, but in a specific and
system-dependent way rather than in a generic manner. Our results were
obtained in a modified two-particle kicked top, i.e. a kicked top involving
explicitly two coupled angular momenta, one for each of the two entangled
particles.\ This modified kicked top is different from the standard kicked
top, which is well-known to be one of the prototypical systems of quantum
chaos \cite{book}. Indeed, the standard kicked top involves a single angular
momentum kicked by an external force whereas in the model we employed, an
angular momentum is kicked by an interaction with the second particle,
inducing a change in both particles' angular momentum. Accordingly, the
classical dynamics of our modified kicked top is richer than the dynamics of
the standard top. Notwithstanding, it can be rigorously shown (see below)
that the standard kicked top is a limiting case of the modified kicked top
we employed in our previous studies.

In the present paper, we will study the entanglement generation as a
function of the underlying classical dynamics of the standard kicked top.
Although the standard top involves a single angular momentum, this angular
momentum can be considered as a composite angular momentum resulting from
the entanglement of several elementary angular momenta (eg, qubits).
Previous results \cite{ghose04,stamatiou07,ghose08} on the standard top employed in this
context suggested that entanglement generation in the quantum top is
correlated with chaos in the classical counterpart, and that in this system
entanglement could be seen as a signature of chaos. These results appear to
conflict the results we obtained with the modified kicked top. The physical
origin of entanglement is surely different as one goes from the modified to
the standard top, but the nature of the quantum-classical correspondence is
not expected to change.\ We therefore investigate in this work the
relationship between entanglement generation and the quantum-classical
correspondence in order to assess to what extent entanglement can be taken
as a signature of chaos.

We will start in Sec.\ II by introducing the standard kicked top, not in the
usual way but as the limiting case of our modified kicked top employed in
earlier works. By doing so we will establish the relationship between these
two models both based on stroboscopic maps. We will then describe the
single standard kicked top as a compound system (Sec.\
III); we will see that the reduced linear entropy (that we will take to be
the marker of the entanglement rate) depends on the averages of the angular
momentum projections. The entanglement
rate is maximized when the sum of these averages is minimized. From the
quantum-classical correspondence viewpoint, the pertinent variable will
consist in obtaining the classical dynamics minimizing this sum. The results
are given in Sec.\ IV, for regular, mixed phase-space and chaotic dynamics.\
We will see that generically, chaos indeed minimizes this sum, but for
appropriately chosen initial states, regular dynamics can entangle more
efficiently, and in a more controlled fashion.\ We will discuss our results
and conclude in Sec.\ V.

\section{The kicked top: an alternative derivation}
We present in this section the link between the Rydberg molecule model, that we employed in our previous studies \cite{epl06,pra06,lasphys10} on entanglement generation and the quantum-classical correspondence, and the well-known kicked top (whose entanglement properties relative to the underlying classical dynamics will be studied in Secs. III-V; readers solely interested in the kicked top results may jump directly to Sec. III). The aim of this section is to show that the standard kicked top can be seen as the limiting case of the Rydberg molecule model when the total angular momentum and one of the two coupled angular momenta become infinitely large.

\subsection{The Rydberg molecule model: torsion and rotation of two coupled
angular momenta}

\subsubsection{Historical Introduction}

Electronic states of atoms or molecules are called Rydberg states, as opposed to valence states, when an outer electron moves far away from the remaining ionic core. These states form  electronic series which converge towards the ionization limit of this outer electron.

The starting point of the quantum analysis of such states was the Quantum
Defect Theory (see e.g. the review article by Seaton \cite{seaton83}), established first for atoms. It was shown that, due to the non zero spatial extension of the ionic core, the levels near the ionization limit follow the hydrogen Rydberg law $E_n=-\mathrm{Ry}/(n+d)^2$, with only a constant (or nearly so) shift $d$ of the principal quantum number $n$, entitled Quantum Defect.
Quantum Defect Theory was extended to Multichannel Quantum Defect Theory (MQDT), for the case that there are several series which converge to nearby states of the ion, and interact strongly. This theory depends only on a small number of parameters, basically one quantum defect per interacting series.
Practically, all is solved with matrices whose size is the number of series, while ``brute force'' methods would in principle try to diagonalize a matrix which contains an infinite number of levels for each series.

This theory was extended to molecules by Fano \cite{fano70,fano75}. There are always many interacting series corresponding to the rotational states of the ionic core. Indeed the slow velocity of the core rotation leads to a splitting of the rotational states of the core which is of the same order of magnitude as the splitting between high lying electronic Rydberg states. The novelty was the implications of the anisotropy of the core. The effect of this anisotropy on the ionic potential decays faster with distance $r$ than the point charge $1/r$ Coulomb potential, at least as $1/r^2$ or $1/r^3$. Fano showed that the key point of the analysis is the existence of a cut off distance $r_0$. Below this distance the motion of the outer electron is tightly bound to the \emph{direction} of the ionic core, above it the two become independent. Many detailed studies have followed on moderately excited Rydberg states of molecules, see e.g. reviews in refs.~\cite{greene85,jungen96}.

\subsubsection{Phase Space: Dimension and Coordinates}

\begin{figure}[tb]
\begin{center}
\includegraphics[angle=270,width=6cm]{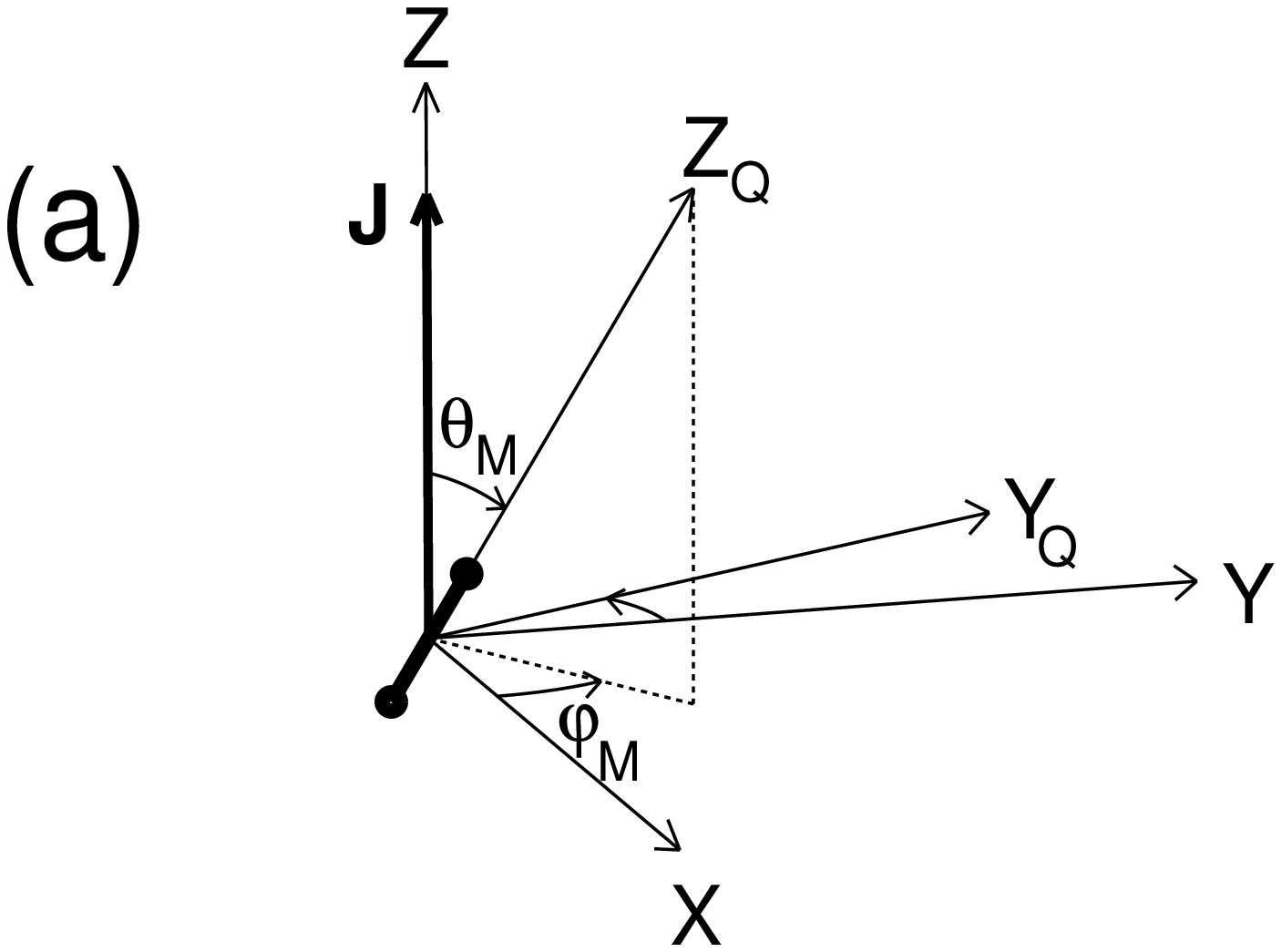}
\end{center}
\begin{center}
\includegraphics[angle=270,width=6cm]{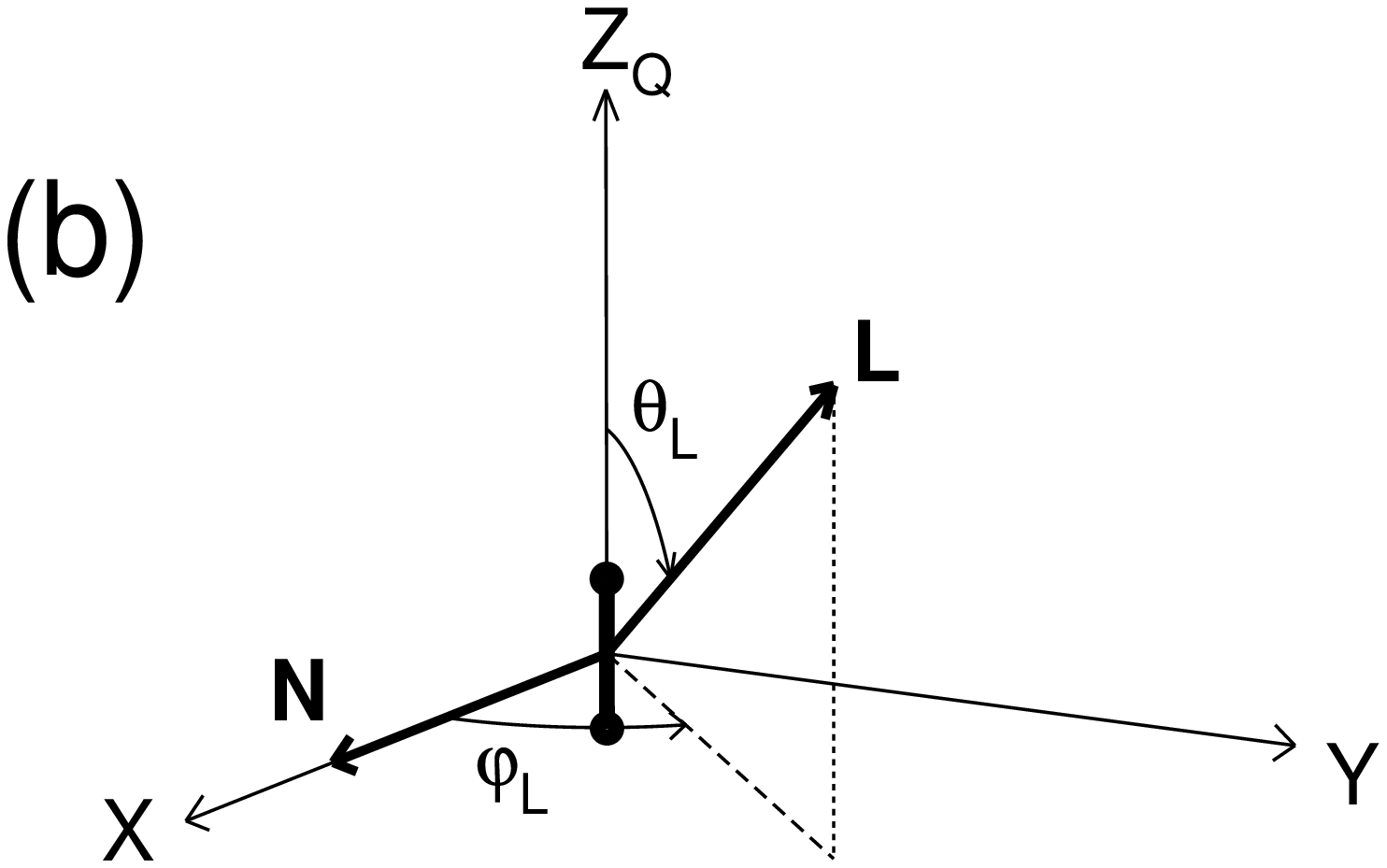}
\end{center}
\caption{(a): The laboratory frame, in which the core (depicted by the the two atomic nuclei) rotates approximately freely. $\unitvec{M}$ is along $OZ_Q$. (b) The collision (or molecular) frame: it is the preceding $OX_QY_QZ_Q$ rotated around $OZ_Q$ so that the new $OX$ is along $\vec{N}$. During the collision $\vec{L}$ rotates around $OZ_Q$ by an angle $\delta \varphi_L$. The ''free'' rotation appears in this frame as a rotation of $\vec{L}$ around the new $OX$. }
\label{figrep}
\end{figure}

For a diatomic molecule this problem is in principle a three body problem, the two ions which constitute the molecular core and the Rydberg electron. After separation of the center of mass motion it depends in configuration space on six parameters, which can be chosen as the coordinates $R, \theta_M, \varphi_M$ of the relative position $\vec{M}$ of the two ions, and the coordinates $r, \theta_e, \varphi_e$ of the Rydberg electron (all in the laboratory frame moving with the center of mass). The two coupled angular momenta we will study here are the angular momentum of the core, $\vec{N}$, associated with the angles $\theta_M, \varphi_M$, and the angular momentum of the Rydberg electron $\vec{L}$, associated with the angles $\theta_e, \varphi_e$ (see Fig. \ref{figrep}).

The problem for our purpose is simplified by two approximations \cite{lombardi88,dietz04}, $R$ and the modulus $L$ of $\vec{L}$ are kept constants. The first amounts to neglecting the vibrational motion of the core, and is frequently fairly well valid, since vibrational quanta are much greater than rotational quanta of the core. The second supposes that the isotropic part of the non Coulomb short range part of the Rydberg electron - ionic core interaction potential is much greater than its non isotropic part. Its validity is often acceptable.
With these two approximations, the dimension of the classical phase space is decreased from $2*6$ by $2*2$ to $8$. Taking into account the conservation of the total angular momentum $J$ and its projection $J_Z$ on the laboratory axis, decreases further this dimension by $2*2$, i.e. the dimension is equal to four. Two classical phase space coordinates can be chosen as $r$ and its conjugate momentum $p_r$. For reasons which come from the quantum treatment below, the other two can be chosen either \emph{in the collision frame} as the projection $L_{Z_Q}$ of $L$ onto the $OZ_Q=\unitvec{M}$ axis and its conjugate angle, the angle of the projection of $L$ into the $OX_QY_Q$ plane with the $OX_Q$ axis (known as $\pi/2$ plus the ascending node in classical mechanics treatises \cite{whittaker60}), or \emph{in the laboratory frame} as the modulus $N$ of the momentum of the ionic core and its associated angle $\varphi_N$, which precise value is given in \cite[Eq.~(A.11)]{dietz04}.

\subsubsection{Quantum solution: MQDT as a quantum map}

The usual manner of obtaining the solutions with MQDT is described in the Appendix. It involves introducing two different angular bases, the collision basis and the free rotation basis. The former gives the correct physical description when the electron is near the molecular core (and therefore collides) whereas the latter corresponds to a large radial distance between the electron and the core (which then rotates freely in the laboratory frame). For our present purposes, it is however more meaningful to envision the quantum problem as a quantum map. Indeed the quantization condition given by Eq.~(\ref{eq:EqClosed}) obtained from MQDT can be rewritten as \cite{leyvraz00,dietz04}
\begin{equation}\label{eq:EtE=1}
    \transpose{\matE} \matE \vert A_\Lambda\rangle = \Id \vert A_\Lambda \rangle,
\end{equation}
with the complex symmetric matrix $\matE$ defined by
\begin{equation}\label{eq:E=nuUmu}
    \matE = \exp(\rmi\pi\matr{\nu})\ \matr{U} \exp(\rmi\pi\matr{\mu})
\end{equation}
where $\matr{\nu}$ and $\matr{\mu}$ are diagonal matrices with diagonal elements $\nu_N$ and $\mu_\Lambda$ and $\matr{U}$ is a unitary matrix given right below.
This equation has the following interpretation \cite{dietz04}. $\vert A_\Lambda \rangle$ is the set of angular coefficients in the ``collision basis" (\ref{eq:collisionwf}) of the wavefunction at the perigee of the trajectory. Eq.~(\ref{eq:EtE=1}) means that at the quantized energy this wavefunction goes back onto itself when applying in order (from right to left) the following operations:
\begin{enumerate}
  \item a diagonal matrix with element $\exp(\rmi\pi\mu_\Lambda) = \exp(-\rmi\pi k/(4 \pi L)\,\Lambda^2)$
      (see Eq.~\ref{eq:mulambda}), i.e. (half) a \emph{quadratic} torque in the collision frame with an angle of $\pi$, a strength of $k/(4\pi L)$ and an operator $L_{Z_Q}^2$, which brings the wavefunction from perigee to outgoing $r=r_0$,
  \item a matrix $\matr{U}_{N\Lambda}$ which transforms from the ``collision basis" (\ref{eq:collisionwf}) labeled by $\Lambda$  to the ``free rotation basis" (\ref{eq:coulombwf}) labeled by $N$,
  \item a diagonal matrix with element $\exp(\rmi\pi\nu_N) \sim \exp(-\rmi\pi\frac{T_e}{T_c} N)$, (plus a constant phase) to first order in $N$ \cite{dietz04}, where $T_e$ and $T_c$ are respectively the average periods of the electron orbit and of the free rotation of the core (see also the derivation below Eq.~(\ref{eq:nuNlin})), i.e. \emph{approximately} (half) a \emph{linear} rotation with angle $\pi$, a strength $T_e/T_c$ and an operator $N$, which brings the wavefunction onto its apogee,
\end{enumerate}
and then applying the same operators in reverse order to bring the wavefunction back to its perigee.
This means that at quantized energies the wavefunction is invariant (not merely unitary as at other energies) under the action of a matrix $\transpose{\matE} \matE$ which is a Quantum Poincar\'e Map \cite{dietz04} in angular space from perigee to perigee.

Notice then that by multiplying Eq.~(\ref{eq:EtE=1}) to the left by $\exp(\rmi\pi\matr{\mu})$ and reorganizing the result gives the equivalent equation
\begin{equation}\label{eq:OutgoingMap}
    \exp(\rmi 2\pi\matr{\mu}) \transpose{\matr{U}} \exp(\rmi 2\pi\matr{\nu}) \matr{U} \vert A^\rmo_\Lambda\rangle = \Id \vert A^\rmo_\Lambda \rangle,
\end{equation}
with
\begin{equation}\label{eq:OutGoingwf}
    A^\rmo = \exp(\rmi\pi\matr{\mu}) A
\end{equation}
which means with the same reasoning that the $A^\rmo_\Lambda$ are the coefficients of the angular wavefunction at the outgoing $r=r_0$ position in the collision basis, and that this wavefunction at quantized energy $E$ is invariant by the quantum Poincar\'{e} map given by Eq.~(\ref{eq:OutgoingMap}).
This is the equation we will employ to derive the kicked top model.


\subsubsection{Classical model: the stroboscopic map}

The preceding quantum theory has a classical counterpart \cite{lombardi88} which is a succession of two different motions (see Fig.~\ref{figrep}(b)):
\begin{itemize}
  \item When the electron is far from the core, their motions are not coupled, due to the rotational invariance of the Coulomb potential. The electron freely rotates around its angular momentum $\vec{L}$ fixed in space, and the core directed along $\vec{M}$ rotates around its angular momentum $\vec{N}$ perpendicular to it. When seen in the molecular reference frame, which $OZ_Q$ axis is $\unitvec{M}$, the rotation of $\vec{L}$ is in retrograde direction. This is the reason of the ``anomalous commutation rules of momentum in molecular basis" \cite{landau77}. The angle of rotation of this apparent motion at angular velocity $\omega_c=\partial(BN^2)/\partial N=2 B N$ during the time of an orbit of the Rydberg electron, which has angular velocity $\omega_e=\partial(-1/(2\nu_N^2))/\partial \nu_N = 1/\nu_N^3$ is $\delta\varphi_L=-2\pi\,T_e/T_c=-2\pi\times 2BN\nu_N^3$.
      This is \emph{approximately only} a uniform rotation of the average angle $\delta\varphi_L=-2\pi\times 2BJ\nu_J^3$ when $L \ll J$, so that $N$, which varies between $J-L$ and $J+L$ varies only slightly.
  \item During the collision the motions are coupled. $\vec{L}$ rotates around $\unitvec{M}$ by an angle $\delta\varphi_L=-2\pi\partial\mu_\Lambda/\partial\Lambda=k\Lambda/L$.
      Since the total angular momentum $\vec{J}=\vec{N}+\vec{L}$ (which are both well defined in classical mechanics during the collision) is conserved, this relative motion of $\vec{L}$ around $\unitvec{M}$ entails a ``recoil" of the molecular reference frame whose precise value was computed in \cite{lombardi88}.
\end{itemize}

\subsection{The kicked top derived from the Rydberg molecule model}

\subsubsection{Evolution operator and wavefunctions}

The preceding model is similar to the standard kicked top in that it displays a succession of quadratic torques around $\unitvec{M}$ separated by approximately pure rotations for $\vec{L}$ around the perpendicular axis $\vec{N}$ . But the non exactness of the pure rotation and the recoil motion (modifying $\vec{N}$) during the collision step entails extra complexities (and extra interests arising from the explicit coupling of the angular momenta $\vec{L}$ and $\vec{N}$).
We now establish that for $L \ll J$, our Rydberg molecule model becomes a standard kicked top. Hence the kicked top can be considered to be a special limit of MQDT.

To this end we first compute $\nu_N$ (see Eq.~(\ref{eq:nuN})) for a generic value of $N$ as a function of the middle $\nu_J$, for $N=J$, which is related to total energy $E$ by $E=B J(J+1) -1/(2\nu_J^2)$. $\nu_J$ can be considered as another measure of total energy, in fact a linearized function of energy with average unity spacing in each $N$ series (``unfolded" in the language of Random Matrix Theory \cite{mehta04}).
\begin{eqnarray}\label{eq:nuNlin}
  \nu_N &=& \frac{\nu_J}{\sqrt{1+2B\nu_J^2(N-J)(N+J+1)}}\nonumber \\
  &=& \frac{\nu_J}{\sqrt{1+\frac{T_e}{T_c} \frac{N-J}{\nu_J}\frac{N+J+1}{J}}}\nonumber \\
  &\to& \nu_J +\frac{T_e}{T_c}(J-N)
\end{eqnarray}
with the average ratio of electron to core periods being given by
\begin{equation}\label{eq:TeTc}
   \frac{T_e}{T_c}=\frac{\omega_{N=J}}{\omega_e}
   = \frac{\partial E/\partial J}{\partial E/\partial \nu_J}
   = 2 B J\ \nu_J^3.
\end{equation}
The limit in Eq.~(\ref{eq:nuNlin}) supposes that $\frac{T_e}{T_c}$ and $L$ (thus approximately $N-J$) remain constant, while $\nu_J\to\infty$. Thus according to Eq.~(\ref{eq:TeTc}) we must have $2 B J\to 0$, which we suppose satisfied by $B\to 0$ and $J\to\infty$, thus also $N\sim J\to\infty$, and $(N+J+1)/J\to 2$. The important result is the linearity of $\nu_N$ with respect to $J-N=M_L$, which varies between $-L$ and $+L$. Indeed in this limit $M_L$ is, to first order in $L/J$, equal to the projection of $\vec L$ onto $\vec J$ (obtained by developing $N^2=\vert \vec{J}-\vec{L} \vert^2 = J^2+L^2-2 \, \vec{L} \cdot \vec{J}$). $\vec J$ being invariant can be without loss of generality taken as laboratory $OZ$ axis, so that $M_L$ is the projection of $L$ onto the laboratory $OZ$ axis.

Using the asymptotic formula for Clebsch Gordan coefficients (with the conventions of Edmonds \cite[A2.1]{edmonds74}) when two angular momenta go to infinity, the remaining being finite, the elements of the transformation matrix become
\begin{eqnarray}\label{eq:dL}
    U_{M_L \Lambda}&=& (-1)^{L+\Lambda}\mathfrak{D}^L_{-M_L -\Lambda}(0,\pi/2,0)\nonumber\\
     &=& (-1)^{L+\Lambda} \mathfrak{D}^L_{\Lambda M_L}(0,\pi/2,0),\label{info2}
\end{eqnarray}
where the last expression makes use of \cite[eqs.~(4.2.5)-(4.2.6)]{edmonds74}.
Here $\mathfrak{D}$ is the ``standard" \cite{fanoracah59} or ``passive" \cite{edmonds74,landau77} rotation matrix, i.e. when rotating the reference frame with Euler angles $\alpha,\beta,\gamma$ while keeping fixed the quantum system:
\begin{equation}\label{eq:DL}
   \mathfrak{D}^L_{m m^\prime} = \langle L m\vert
   \rme^{+\rmi\gamma L_z} \rme^{+\rmi\beta L_y}\rme^{+\rmi\alpha L_z}
   \vert L m^\prime \rangle
\end{equation}
Explicit expressions are given in these textbooks \cite{fanoracah59,edmonds74,landau77}, and we have written fast and accurate recursive programs programs for large momenta. Notice the $+$ sign, and angle ordering opposites to the ``active" point of view \cite{messiah64} (rotating the spin in a fixed reference frame), which is more common in kicked top works. The passive point of view is more ``natural" in molecular works \cite{landau77}.

Inserting the expressions  (\ref{eq:nuNlin}) and (\ref{info2}) into Eq.~(\ref{eq:OutgoingMap}) and moving the common factor $\exp(2\rmi\pi\nu_J)$ to the right hand side  the quantization condition
\begin{equation}\label{eq:KSmap}
    \sum_\Lambda
    \rme^{-\rmi k \frac{{\Lambda^\prime}^2}{2L}}
    \mathfrak{D}^L_{M_L \Lambda^\prime}(0,-\pi/2,0)
    \rme^{2\rmi\pi \frac{T_e}{T_c} M_L}
    \mathfrak{D}^L_{\Lambda M_L}(0,\pi/2,0)
    \vert \tilde{A}^\rmo_\Lambda\rangle
    = \rme^{- 2\rmi\pi \nu_J}
    \delta_{\Lambda \Lambda^\prime}
    \vert \tilde{A}^\rmo_\Lambda\rangle
\end{equation}
where we have defined the elements of $\vert \tilde{A}^\rmo_\Lambda\rangle$ by
\begin{equation}\label{eq:OutGoingRydwf}
    \tilde{A}^\rmo = (-1)^{J-\Lambda} A^\rmo .
\end{equation}
This change of sign takes into account that the molecular wavefunction contains a core part in addition to the Rydberg electron part (Eq.~\ref{eq:collisionwf}). We want an equation for the Rydberg electron only, and for $L \ll J$ the core parts for different $\Lambda$ differ only by this sign.
To interpret the quantization condition, it can be noted that for a ``passive" rotation, the transformation
\begin{equation}\label{eq:BM}
    \tilde{B}^\rmo_{M_L} = \sum_\Lambda \tilde{A}^\rmo_\Lambda \mathfrak{D}^L_{\Lambda M_L}(0,\pi/2,0)
\end{equation}
yields the coefficients of the outgoing wavefunction in a frame rotated around $OY$ by $\pi/2$ from the $OZ$ axis (see e.g. \cite[Eq.~(58.7)]{landau77}), that is the $OX$ axis. Alternatively notice that Eq.~(\ref{eq:KSmap}) is an eigen system of equations with eigenvalue $\exp(-2\rmi\pi \nu_J)$ for a unitary map consisting from right to left: a frame rotation from $OZ$ to $OX$, a pure rotation with parameter $\frac{T_e}{T_c}$ along the new axis, back to original axis system (giving overall a free rotation along $OX$), and finally a quadratic torque with parameter $k$ along the original $OZ$ axis. Therefore this is exactly the evolution equation from kick to kick for the standard kicked top \cite{haake}. Notice only the plus sign in the pure rotation term, which appears as the consequence of the apparent rotation in ``wrong" sense when viewed in the molecular frame. Conversely one can say that this sign is correct in the laboratory frame, and that two other minus signs are a consequence of the map being described in the molecular rotating axis system.

\section{Entanglement in a single kicked top}

\subsection{The kicked top as a compound system}

A kicked angular momentum $J$ of the standard top can be considered as
resulting from the composition of $2J$ spin-1/2 subsystems so that $\mathbf{J%
}=\sum_{n=1}^{2J}\mathbf{s}^{n}$ \cite{ghose04}. The spins $\mathbf{s}^{n}$
must be in a state symmetric by permutation in order to generate the Hilbert
subspace for the states $\left\vert JM\right\rangle $ of the kicked top.\
Then the average for an individual spin $\mathbf{s}^{n}$ projection on the
axis $i=x,y,z$ is identical for each $n$ and related to the averages $%
\left\langle J_{i}\right\rangle $ of the total angular momentum through
\begin{equation}
\left\langle s_{i}^{n}\right\rangle =\frac{\left\langle J_{i}\right\rangle }{%
2J}.  \label{r0}
\end{equation}%
We will be interested in the entanglement between an individual spin $%
\mathbf{s}$ and the remaining subsystem containing the $2J-1$ other spins.
The average dynamics for $\mathbf{s}$ are obtained from the density matrix
for the kicked top $\rho (t)$ by taking the partial trace over the remaining
spins.\ Note however that this step need not be done explicitly given that
any spin-1/2 density matrix can be written as%
\begin{equation}
\rho _{s}(t)=\frac{1}{2}+2\sum_{i}\left\langle s_{i}\right\rangle s_{i},
\label{r1}
\end{equation}%
so that the expansion coefficients are actually encoded in the averages. This
is a considerable simplification relative to the modified kicked top.

\subsection{Entanglement generation and the quantum-classical correspondence}

\subsubsection{Linear entropy}

We will quantify entanglement by computing the linear entropy $S_{2}(t)$
associated with the reduced density matrix defined by%
\begin{equation}
S_{2}(t)=1-\text{Tr}\rho_{s}^{2}(t).  \label{r2}
\end{equation}
For a pure state $\rho_{s}^{2}=\rho_{s}$ and $S_{2}$ vanishes, whereas for a
maximally mixed qubit $S_{2}=1/2.$ By plugging Eqs. (\ref{r0})-(\ref{r1})
into Eq.~(\ref{r2}), we have%
\begin{equation}
S_{2}(t)=\frac{1}{2}-\frac{1}{2J^{2}}\left( \left\langle J_{x}\right\rangle
^{2}+\left\langle J_{y}\right\rangle ^{2}+\left\langle J_{z}\right\rangle
^{2}\right) .  \label{r5}
\end{equation}
Hence entanglement depends on the averages of the kicked top angular
momentum projections; entanglement is maximal when all these averages vanish.

\subsubsection{Coherent states}

Angular momentum coherent states \cite{coherent} are the most suitable
choice in order to investigate the quantum-classical correspondence in the
kicked top, and have consequently been employed from the early works onward
\cite{book}. These coherent states, given in terms of the angular momentum eigenstates by%
\begin{equation}
\left\vert \theta ,\phi \right\rangle =(1+\tan ^{2}\frac{\theta }{2}%
)^{-J}\sum_{M=-J}^{J}\left( \begin{array}{c}
                              2J \\
                              J-M
                            \end{array}
\right) ^{1/2}\left( \tan \frac{%
\theta }{2}e^{i\phi }\right) ^{J-M}\left\vert JM\right\rangle
\end{equation}%
are localized on the sphere, and in the present context they present the
additional advantage of yielding an initial product state. Indeed, from the
property%
\begin{equation}
\left\langle \theta ,\phi \right\vert J_{i}\left\vert \theta ,\phi
\right\rangle =J_{i},
\end{equation}%
if $\rho (t=0)=\left\vert \theta _{0},\phi _{0}\right\rangle \left\langle
\theta _{0},\phi _{0}\right\vert $ we then have $\left\langle J
_{i}(t=0)\right\rangle=J_{i}$ resulting in $S_{2}(t=0)=0$.

The most straightforward way of representing a coherent state and its
ensuing evolution on the sphere is through the use of a Husimi distribution,
which is precisely defined as the coherent state representation of the
density matrix. For the standard kicked top in state $\left\vert
\psi\right\rangle $ this is simply given by%
\begin{equation}
h(\theta,\phi)=\left\vert \left\langle \theta,\phi\right. \left\vert
\psi\right\rangle \right\vert ^{2}.
\end{equation}
When $\left\vert \psi\right\rangle $ is itself a coherent state $\left\vert
\psi\right\rangle \equiv\left\vert \theta_{0},\phi_{0}\right\rangle $ the
Husimi distribution is given by the overlap of the coherent states%
\begin{equation}
\left\vert \left\langle \theta,\phi\right. \left\vert \theta_{0},\phi
_{0}\right\rangle \right\vert ^{2}=\cos^{4J}(\frac{\chi(\theta\phi,\theta
_{0}\phi_{0})}{2})  \label{r10}
\end{equation}
where $\chi$ is the angle between the directions $\left( \theta,\phi\right) $
and $\left( \theta_{0},\phi_{0}\right) $. This gives a distribution
localized on $\left( \theta_{0},\phi_{0}\right) $ with an angular spread
inversely proportional to $J$.

In the rest of this work we will choose an initial coherent state centered on $\left(
\theta_{0},\phi_{0}\right) $ at $t=0$ and monitor the entanglement
generation by computing $S_{2}(t)$ that depends on the averages of the
angular momentum projections through Eq.~(\ref{r5}).

\begin{figure}[tb]
\includegraphics[height=4cm]{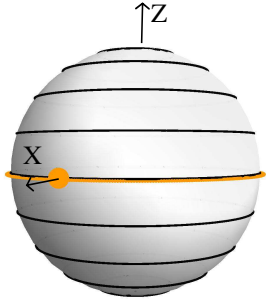}
\caption{Surface of section at ''resonance'' (see text).
 }
\label{figx1}
\end{figure}

\subsubsection{Classical distributions and averages}

Several aspects of the quantum-classical correspondence in the kicked top
have been investigated \cite{haake,fox94}. Here we will only be
interested in comparing the quantum averages $\left\langle
J_{i}(t)\right\rangle $ to the corresponding classical averages $\bar{J}%
_{i}(t)$, when the initial quantum distribution is the coherent state $%
\left\vert \theta _{0},\phi _{0}\right\rangle $ and its classical
counterpart is a distribution of particles centered on $\left( \theta
_{0},\phi _{0}\right) $ and distributed on the sphere according to the right
hand-side of Eq.~(\ref{r10}). For very short-time scales the classical and
quantum averages are expected to be the same, though for longer times in
typical cases at best a similar qualitative behavior can be obtained when
the statistical averages are averaged out on relevant time scales in order
to smooth out interference effects (such as quasiperiodic quantum revivals
when the underlying classical dynamics is regular or the random
superpositions when the underlying dynamics is chaotic).

\begin{figure}[tb]
\includegraphics[height=4cm]{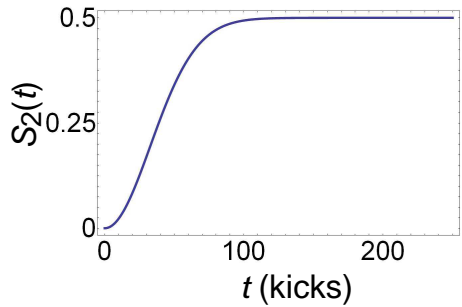}
\caption{The linear entropy as a function of time when the initial state is
a coherent state centered on $(\pi/2,0)$ (resonant case, $k=0.1$). }
\label{QresoK01}
\end{figure}

Our interest in the quantum-classical correspondence in the present context
is to assess whether the configurations (kick strength, relative rotational
and torsional frequencies, initial state position) leading to efficient
entanglement generation in the quantum top can be related to different
dynamical regimes of the classical top. For these purposes, a numerical
comparison of the linear entropy [Eq.~(\ref{r5})] and of the corresponding
classical expression%
\begin{equation}
C_{2}(t)=\frac{1}{2}-\frac{1}{2J^{2}}\left( \bar{J}_{x}^{2}+\bar{J}_{y}^{2}+%
\bar{J}_{z}^{2}\right)  \label{r15}
\end{equation}
is sufficient. We can further expect that the similar behavior of $S_{2}$
and $C_{2}$ is due to the similar behavior of the classical and quantum
angular momentum averages. Contrarily to the situation of genuine 2-particle
systems (in particular in our modified top), here $C_{2}$ is totally unrelated to a measure of
the non-separability of phase-space distributions that could classically play the role
of $S_2$ as a marker of classical mixtures \cite{classent}. Intuitively, it can be anticipated that starting from a localized
state on the sphere, chaotic dynamics will tend to scatter the initial
distribution over the entire sphere, leading to vanishing or small $\bar {J}%
_{i}$ for the three axes.\ This will lead to a maximization of $C_{2}(t)$,
and should correspond to maximal entanglement generation.

\begin{figure}[tb]
\includegraphics[height=6cm]{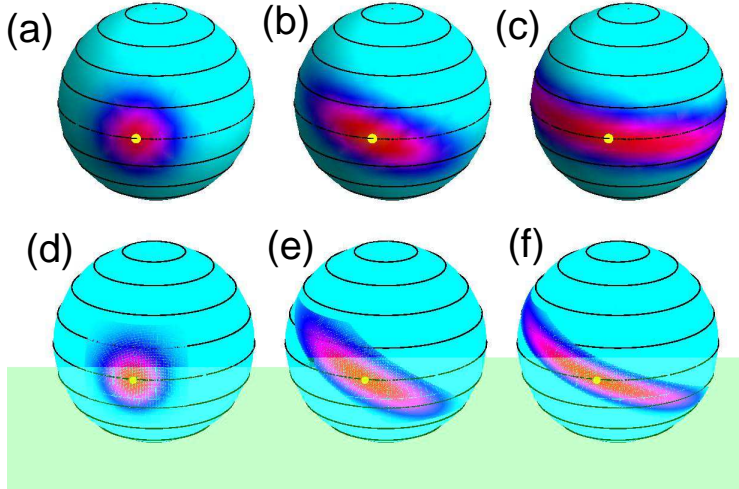}
\caption{(Color online). Evolution of the quantum Husimi distribution (top row) and the
analog classical distribution (bottom row). The plots show the short-time
evolution of the distributions for times corresponding to the rise in $%
S_{2}(t)$ seen in Fig.~\ref{QresoK01} (resonant case, $k=0.1$). The
initial coherent state centered on the yellow dot in (a) and its classical
counterpart in (c) is subjected to the torsional motion resulting in a
spread of the distribution on a strip on both sides of the equator (the
coloring reflects the intensity of the distribution).}
\label{HusimClass}
\end{figure}

\section{Results}

We will compute numerical results for the quantum and classical tops with $%
J=10$.\ This moderate value of the angular momentum is sufficiently low as
to envisage the possibility of an experimental realization while already
displaying the main features of the quantum-classical correspondence. We keep to the
conventions introduced in Sec. II (i.e. free rotation around the $OX$ axis
and torsion along the $OZ$ axis).

\subsection{Regularity at ''resonance''}

In the Rydberg molecule, resonance refers to the electron period $T_e$ being an integer multiple
of half the core period $T_c$ (half is due to Kr\"onig's symmetry of the core \cite{HerzbergI50}). This situation has observable consequences, appearing as clear zones in the spectrum \cite{lombardi88}, and
achieves high entanglement generation when the dynamics is regular \cite{pra06}.
In the standard kicked top limit, the free rotation becomes trivial: the classical
dynamics is always regular (irrespective of $k$) and constrained to remain on the initial circle,
while increasing $k$ leads to an arbitrary separation between successive
points on the circle.  The corresponding surface of section is displayed in Fig.~\ref{figx1}.

Let us take $k=0.1$ and an initial distribution centered in $\left( \theta _{0},\phi
_{0}\right) =\left( \pi /2,0\right) $ at the intersection of the $x$ axis
with the sphere (orange dot in Fig.~\ref{figx1}). The entanglement rate is shown in Fig.~\ref{QresoK01}.\ We
see that the linear entropy rises slowly and monotonically until it reaches
its maximal value of $1/2$. The classical quantity $C_{2}(t)$ follows
strictly the same behavior. This behavior is due to the effect of the
torsion on the distribution: in Fig.~\ref{HusimClass} we have displayed the
short-time evolution of the quantum (upper plots) and classical
distributions. The initial coherent state, shown in Fig.~\ref{HusimClass}
moves to the left on the upper half of the sphere and to the right on the
lower half of the sphere. This ensures that both $\left\langle
J_{y}(t)\right\rangle $ and $\left\langle J_{z}(t)\right\rangle $ keep their
initial value of $0$. $\left\langle J_{x}(t)\right\rangle $ on the other
hand evolves from initial value $\left\langle J_{x}(t=0)\right\rangle =J$ to
$\left\langle J_{x}(t)\right\rangle =0$ as the distribution stretches and
encircles the sphere along the equator.

\begin{figure}[tb]
\includegraphics[height=8cm]{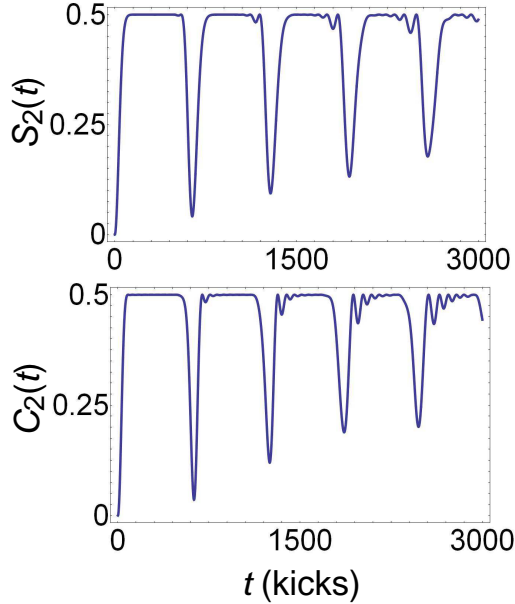}
\caption{Top: Evolution of the linear entropy shown in Fig.~\ref{QresoK01}
for longer times. Bottom: the
analog quantity $C_2 (t)$ for the corresponding classical system shows the
same behavior, due to quantum-classical correspondence for the average of
the projection $J_x$. }
\label{QCresolongK01}
\end{figure}

This feature is readily understood by looking at the evolution of an initial
classical distribution corresponding to the coherent state (bottom row of Fig.~\ref{HusimClass}): the particles far from the center of the distribution
spread faster than those near the center and the first ones reach the
opposite side of the sphere while the latter are still close to $\left(
\theta _{0},\phi _{0}\right) $. After a few kicks the distribution becomes
approximately a uniform strip around the equator during which time $\bar{J}%
_{x}\simeq 0$ and $C_{2}\simeq 1/2$. For longer times the distribution
relocalizes on the opposite side $\left( \pi /2,\pi \right) $ (this is a
purely classical effect) with $\bar{J}_{x}$ almost equal to $-J$ and spreads
again.\ The corresponding behavior of $S_{2}(t)$ for longer times is shown
in Fig.~\ref{QCresolongK01}, along with $C_{2}(t)$.

\begin{figure}[tb]
\includegraphics[height=4cm]{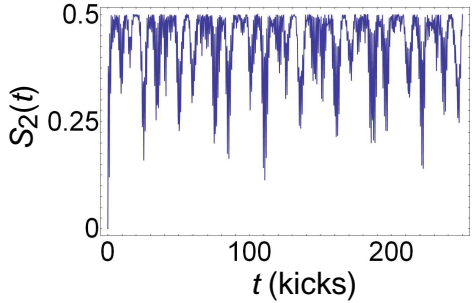}
\caption{Top: $S_2 (t)$ for the ''resonant'' case and $k=10$ (the initial state
is the same coherent state as in Fig.~\ref{QresoK01}). }
\label{QresoK10}
\end{figure}

Note that the time averaged entanglement rate before the first
relocalization is extremely high, $S_{2}\approx 0.46$ (for longer times
partial relocalizations proliferate and the average decreases to $%
S_{2}\approx 0.43$). It is interesting to compare with the case $k=10$. The
entanglement evolution is shown in Fig.~\ref{QresoK10}; $S_{2}(t)$ reaches
the maximal value of $1/2$ in only a couple of kicks, but periodically drops
to significantly lower values (the time average is $S_{2}\approx 0.43$
identical to the $k=0.1$ case).

\subsection{Mixed Phase-Space}

\begin{figure}[tb]
\includegraphics[height=8cm]{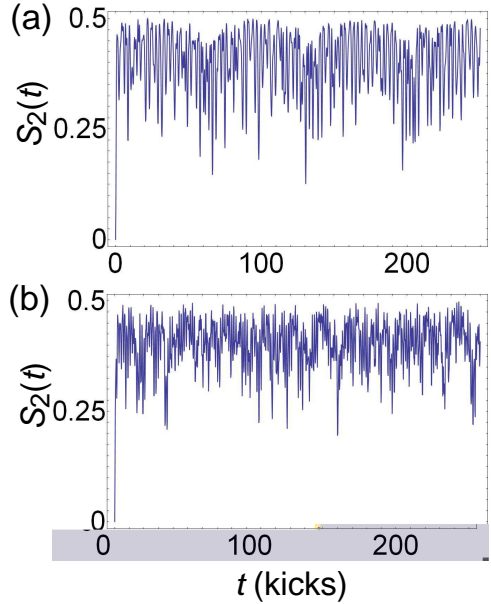}
\caption{$S_2 (t)$ for the case $T_e=0.95$, $k=5$ when the initial coherent
state is chosen to lie in the chaotic sea [(a)] or in the central regular
region [(b)], on the points shown on the surface of section in Fig.~\ref{ClassDisMix}. }
\label{Qmix}
\end{figure}

\begin{figure}[tb]
\includegraphics[height=7.5cm]{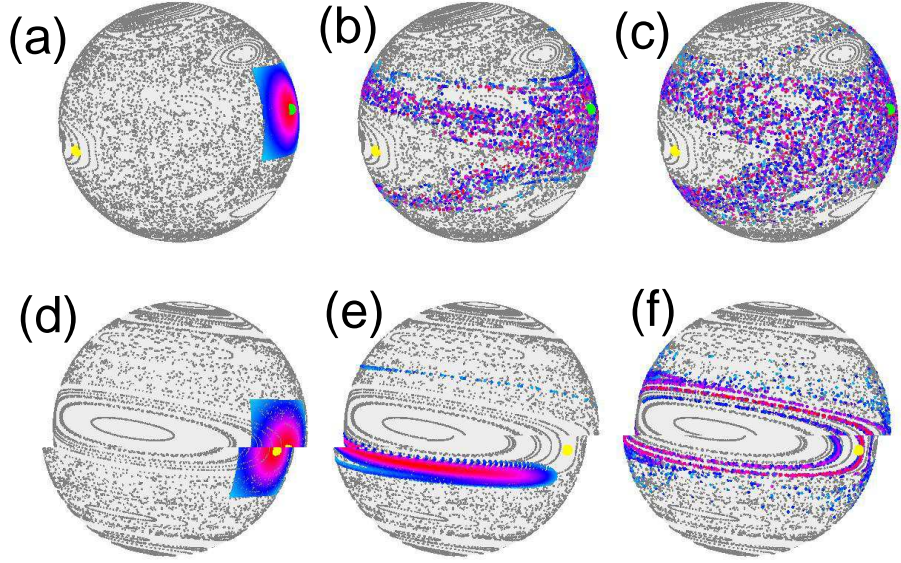}
\caption{(a)-(c): Evolution of a classical distribution for the case $%
T_e=0.95$, $k=5$ when the initial distribution is the analog of the coherent
state centered on the green dot lying in the chaotic sea (whose linear
entropy was plotted in Fig.~\protect\ref{Qmix}(a)). (a) shows the initial
situation, (b) the evolution after 15 kicks, (c) the evolution after 75
kicks. (d)-(f): Same as above with the initial distribution being the analog
of the coherent state centered on the yellow dot lying in the regular region
(the linear entropy was plotted in Fig.~\protect\ref{Qmix}(b)). (d) shows
the initial situation, (e) the evolution after 4 kicks, (f) the evolution after 75 kicks (most of the evolution follows clockwise an elliptical motion and returns near the initial point). }
\label{ClassDisMix}
\end{figure}

Let us now investigate the entanglement generation for the mixed phase situation $T_{e}=0.95$ and $k=5$ (see the classical surfaces of section in Fig.~\ref{ClassDisMix}). We first take an initial state lying in the \emph{chaotic sea} (Fig.~\ref{ClassDisMix}, top panel). The evolution of $S_{2}(t)$ is shown in Fig.~\ref{Qmix}(a). The linear entropy reaches its maximal value of $1/2$ after only a few kicks, but strong dips keep the time averaged value over the first 1000 kicks to $S_{2}\approx0.40$. The evolution of the corresponding classical distribution is shown in Fig.~\ref{ClassDisMix} (a)-(c): the distribution quickly spreads over most of the chaotic sea, that nevertheless only covers a part of the available phase-space.

Let us now take $\left( \theta _{0},\phi _{0}\right) $ centered on a point lying in the large \emph{regular region} as indicated in Fig.~\ref{ClassDisMix} (lower panel). The linear entropy is shown in Fig.~\ref{Qmix}(b). It immediately rises to $S_{2}=0.45$ and then oscillates wildly with a time average of $S_{2}\approx 0.40$. An inspection of the behavior of $\left\langle J_{i}\right\rangle $ shows that $\left\langle J_{y}\right\rangle $ and $\left\langle J_{z}\right\rangle $ display quasi-periodicities typical of quantum revivals in regular systems whereas $\left\langle J_{x}\right\rangle $ follows on average the behavior of the classical average $\bar{J}_{x}$ (see Fig.~\ref{JmoyQC}). The evolution of the classical distribution, shown in Fig.~\ref{ClassDisMix}(d)-(e), indicates that most of the distribution stays within the island of regularity but spreads along the lines in the surface of section. The spreading is not perfectly uniform but the approximate symmetry of the distribution suffices to reduce significantly the averages $\bar{J}_{i}$, which in turn increases $C_{2}(t)$ to values similar to the case when the classical distribution explores the chaotic sea. Note that in these two cases the time-averaged entanglement rate is exactly the same.

\subsection{Chaotic Phase-Space}

For $T_{e}=0.90$ and $k=15$ there are no more islands of stability and the
stroboscopic map is chaotic.\ The entanglement rate is shown in Fig.~\ref{Qchaos}, rising in a couple of kicks to the maximum bound. The time average of the linear entropy is $S_{2}\approx 0.47$. The quantum and classical averages of the angular momentum projections $\left\langle J_{i}(t)\right\rangle $ and $\bar{J}_{i}(t)$ oscillate around zero with a small but far from negligible amplitude (about $\pm 2$), thereby explaining the dips visible in the linear entropy.

\section{Discussion and conclusion}

\begin{figure}[tb]
\includegraphics[height=7cm]{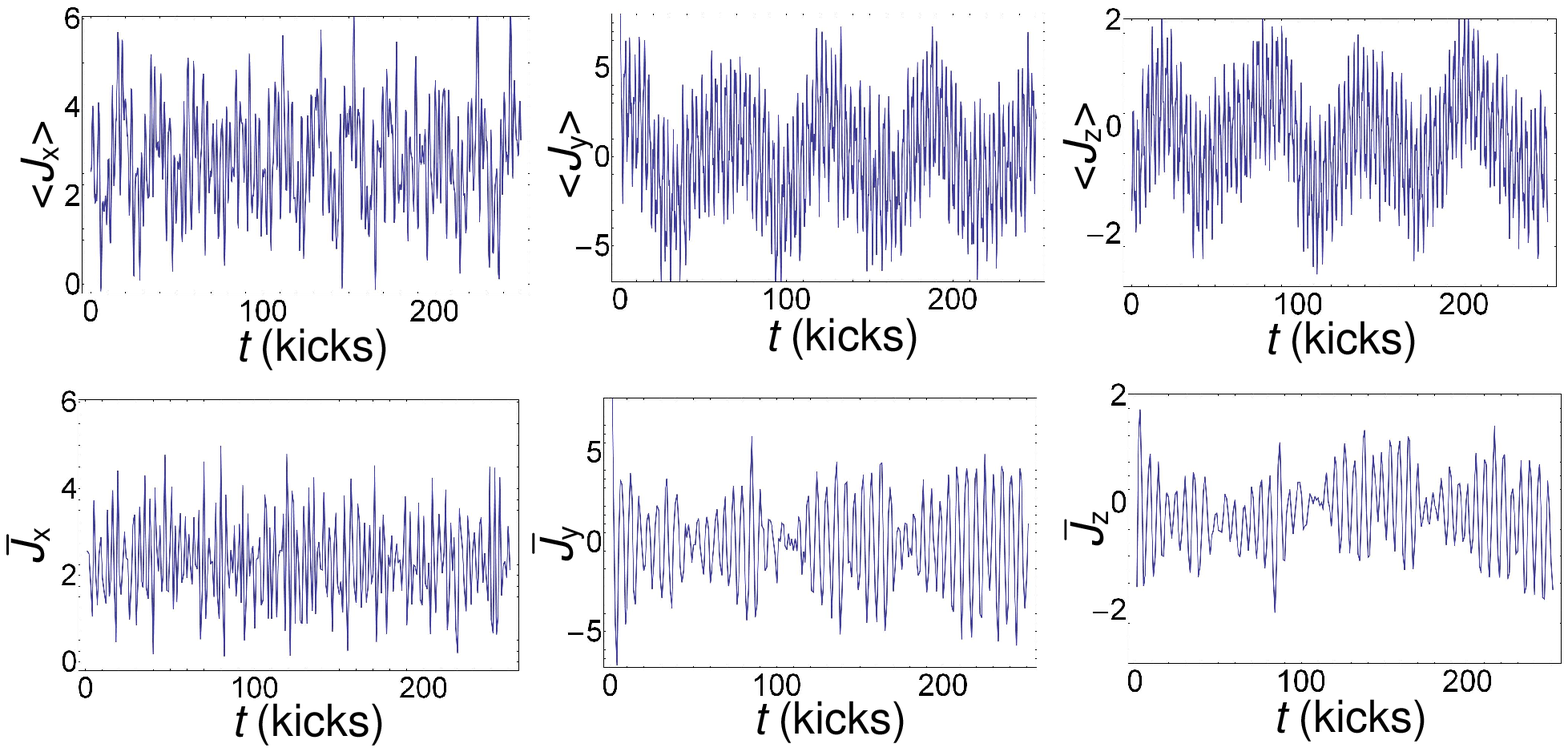}
\caption{The averages of the angular momentum projections
$J_{i}(t)$ for a kicked top with $T_{e}=0.95$ and $k=5$ are
plotted for the quantum top (top row) and the corresponding classical top
(bottom row) when the initial distribution is in the regular region. }
\label{JmoyQC}
\end{figure}
In our previous studies regarding the correspondence between entanglement
generation and the underlying classical dynamics in Rydberg molecule
top \cite{epl06,pra06}, we had found that the linear entropy was correlated
with the classical diffusion in regions of phase-space leading to inelastic
scattering between the angular momentum of each of the two particles. Our
conclusion was that chaotic classical dynamics tended to favor inelastic
scattering, but was by no means a necessary condition, given that regular
dynamics could achieve in certain conditions more efficient inelastic
scattering (resulting in higher entanglement in the quantum system).

\begin{figure}[tb]
\includegraphics[height=4cm]{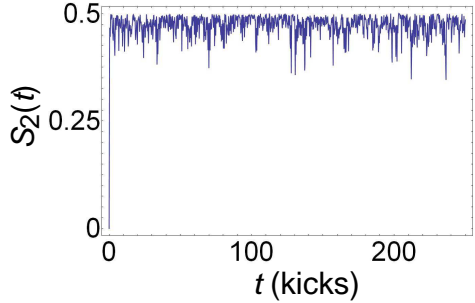}
\caption{$S_{2}(t)$ for the case $T_{e}=0.90$, $k=15$ corresponding to a
classically chaotic phase-space. }
\label{Qchaos}
\end{figure}

In the limiting case yielding the standard kicked top in which the single
angular momentum is regarded as a collection of entangled spins, the
correlation variable is considerably simpler: it only involves the sum of
the angular momentum projection averages.\ The linear entropy given by Eq.~(%
\ref{r5}) has a direct classical counterpart $C_{2}(t)$ given by Eq.~(\ref%
{r15}). Chaotic dynamics tends to maximize $C_{2}(t)$ as the initial
delocalized distribution spreads all over the sphere.\ However we have seen
that regular dynamics can also achieve high values of $C_{2}(t)$ by taking
advantage of the symmetry of the evolving distributions. In the regular
''resonant'' case (Fig.~\ref{QresoK01}) the time averaged entanglement before
the first relocalization is equivalent to that of the entirely chaotic top
of Fig.~\ref{Qchaos}; actually contrarily to the chaotic case, in the
resonant kicked top $C_{2}(t)$ sticks for a period of several hundred kicks
to its maximal value of $0.5$ without any dip.\ Hence the resonant top is an
interesting candidate to control and achieve the highest degree of
entanglement (the main drawback being the slow rise, which may conflict with
decoherence scales in practical applications).

The mixed phase-space top also shows interesting properties: the linear
entropy when the initial state lies in the classically regular or chaotic
regions shows a similar behavior, although the underlying classical dynamics
is radically different, as portrayed in Fig.~\ref{ClassDisMix}. The
initial classical distribution in the chaotic sea  mostly spreads within the
available phase-space region, inducing a rapid rise in $C_{2}(t)$ (which is
not maximal because a large portion of the sphere has regular features). For
the regular initial distribution case the spread takes place essentially
within the regular region.\ The averages shown in Fig.~\ref{JmoyQC} indicate
that $\bar{J}_{y}$ and $\bar{J}_{z}$ are not zero (as they would be for a
uniform distribution) but oscillate instead with a small amplitude for $\bar{%
J}_{z}$ and a larger one for $\bar{J}_{y}$.\ Still, given that $\bar{J}_{x}$
itself oscillates around its small initial value, the resulting $C_{2}(t)$
is large, of the same order of magnitude as when the distribution was
confined to the chaotic sea. Note that the oscillations of $\left\langle
J_{y}\right\rangle $ and $\left\langle J_{z}\right\rangle $ in Fig.~\ref%
{JmoyQC}, that are typical of recoherences and revivals when the
corresponding classical regime is regular cancel out when summed in the
expression of $S_{2}(t)$ and are therefore not visible in the entanglement
rate. Here instead the time dependence of $S_{2}(t)$ in Fig.~\ref{Qmix}(a)
(initial state in the classically chaotic region) displays an aspect reminiscent of
recoherences.

The present findings do not disprove earlier results \cite{ghose04,ghose08}
on chaos and entanglement in the kicked top in the sense that generically
regular classical dynamics will tend to be correlated with lower quantum
entanglement than when the classical dynamics is chaotic. Notwithstanding we
have given explicit illustrations in the kicked top indicating that this
generic behavior is not universal. The reason is that the linear entropy is
correlated through the quantum-classical correspondence to the quantity $%
C_{2}(t)$ introduced above, which in turn depends on the classical
averages.\ $C_{2}(t)$ does not depend simply on the global dynamical
regime but on the details of the initial distribution and on its specific
dynamical evolution. In the examples we have given high values of $C_{2}(t)$
were obtained for distributions evolving through regular dynamics by
appropriately choosing the localization of the initial distribution.\ Note
that this feature can be enhanced by choosing multiply localized initial
distributions (eg a sum of a couple of coherent states each centered on a
different point on the sphere) so that the ensuing regular dynamics
minimizes the averages entering $C_{2}(t)$ or $S_{2}(t)$. We therefore
conclude that as was already seen on other systems
\cite{nemes99,fujisaki04,pra06,zhang08,chung09}, general claims linking chaos
and entanglement in the standard kicked top should be made with care, as they
are not universally valid.

To sum up, we have introduced the standard kicked top as the limiting case
of our two-particle kicked top modeling Rydberg molecules employed in
earlier works and investigated dynamical entanglement in the standard top as
a function of the underlying classical dynamics. By linking the marker of
the entanglement rate to a classical function depending on the angular
momentum projection averages we have seen that the entanglement generation
in the quantum kicked top depends on the specific details of the underlying
classical dynamics rather than depending generically on the global
properties of the classical regime.

\appendix*
\section{Rydberg molecule: Quantum solutions}
The solutions of the quantum problem are obtained by expressing the wavefunction in terms of two different angular bases in order to take into account the short-range phase-shifts (determined in the region near the core) and the asymptotic boundary conditions (for the radial variable going to infinity). The two expressions are matched at some radial value $r=r_0$ for which both decompositions are valid ($r_0$ lying actually near the core).
\begin{description}
  \item[$r<r_0$ Collision or molecular basis.] In this region the Rydberg electron is tightly bound to the core by a strong anisotropic potential, so that we have ordinary molecular diatomic eigenfunctions (Hund's case (b) in molecular nomenclature \cite{HerzbergI50}), where only the total angular momentum $\vec{J}=\vec{N}+\vec{L}$ is conserved (but not separately $\vec{L}$ and $\vec{N}$) so that the good quantum numbers are its modulus $J$, its projection onto the $OZ$ laboratory axis $J_Z=M_J$, and its projection onto the axis of $OZ_Q = \unitvec{M}$ of the core $J_{Z_Q} = L_{Z_Q} = \Lambda$. The last equations suppose that the core itself has no angular momentum along its own axis, i.e. is in a $\Sigma$ state, so that the only angular momentum along the axis $\unitvec{M}$ is that of the Rydberg electron.
      To be definite, we choose the conventions of the original work of Fano~\cite{fano70}, namely
      \begin{itemize}
        \item The (quantum) molecular reference frame is obtained from the laboratory reference frame by a rotation of Euler angles $\varphi_M,\theta_M,0$ (with the ``quantum" convention that the second rotation is around the intermediate $OY$ axis) \cite{fanoracah59,messiah64,edmonds74,landau77}.
        \item there is no extra overall phase factor
      \end{itemize}
      so that the angular part of the basis wavefunctions is
\begin{equation}\label{eq:collisionwf}
    X^{(L J)}_{\Lambda, M_J}(\theta_e^\prime,\phi_e^\prime,\unitvec{M}) =
    Y^L_\Lambda(\theta_e^\prime,\phi_e^\prime)
    \sqrt{\frac{2J+1}{4\pi}}
    \mathfrak{D}^J_{\Lambda M_J}(0,\theta_M,\varphi_M)
\end{equation}
where $\theta_e^\prime, \varphi_e^\prime$ are the polar angles of the Rydberg electron \emph{in the molecular frame}, and where the $\mathfrak{D}$ are given in Eq.~(\ref{eq:DL}).
  \item[$r>r_0$ Coulomb or laboratory or free rotation basis.] In this region the interaction potential is approximated by a pure Coulomb $1/r$ potential which is rotationally invariant. The angular momenta of the core $\vec{N}$ and of the electron $\vec{L}$ are separately invariant in laboratory space, as is their sum the total angular momentum $\vec{J}$. The angular part of the basis wavefunctions is thus obtained by adding the two partial angular wavefunctions with a Clebsch-Gordan coefficient:
\begin{equation}\label{eq:coulombwf}
    \Phi^{(L J)}_{N, M_J}(\theta_e,\phi_e,\unitvec{M}) =
    \sum_{\Lambda M_N}
    \langle L \Lambda, N M_N \vert L N J M_J \rangle
    Y^L_\Lambda(\theta_e,\phi_e)
    \sqrt{\frac{2 N+1}{4\pi}} \mathfrak{D}^N_{0 M_N}(0,\theta_M,\varphi_M)
\end{equation}
the first index of $\mathfrak{D}^N$ is $0$ because the core is in a $\Sigma$ state.
\end{description}

Quantization is obtained by demanding that the wavefunction go to zero at both ends $r\to 0$ and $r\to\infty$.
\begin{description}
  \item[inner region $r<=r_0$.] For a full fledge solution, the radial equation is integrated outwards from $r=0$ to $r=r_0$ with the full anisotropic potential. This is done for each value of the total energy $E$ and each value of the good quantum numbers, which in this range are according to Eq.~(\ref{eq:collisionwf}) $L, J, \Lambda, M_J$. In very serious cases one takes into account that for very short range the interaction is more complex than a mere potential. In any case at $r=r_0$ the radial equation reduces to a second order equation in the isotropic $1/r$ Coulomb potential, which local solutions are known.
      They are a linear combination of the regular and irregular solutions \emph{at origin $r=0$} of this problem as
\begin{equation}\label{eq:collisionradial}
    f_\Lambda(r) = \left( s(E_e ,r) \cos(\pi\mu_\Lambda ) + c(E_e,r) \sin (\pi\mu_\Lambda) \right)
\end{equation}
where $E_e$ is the Rydberg electron energy (\emph{not} the total energy $E$).
It depends on only one parameter $\mu_\Lambda$, which depends itself on $\Lambda, L, J$ (but not $M_J$). It depends also on $E_e$, but very slightly so for highly excited levels, because near $r=r_0$ the attractive Coulomb potential is much greater than the splitting of the rotational energies of the core, which are of the order of the splitting of the higher energy electronic levels which tend to zero at the ionization limit \cite{seaton83,fano70}.
Frequently, instead of computing $\mu_\Lambda$ by radial integration, it is taken as a parameter to be adjusted to experiments.
Due to Kronig's symmetry \cite{HerzbergI50} by reflection on the molecular core axis, it must be a quadratic function of $\Lambda$ \cite{lombardi88}. The simplest case is
\begin{equation}\label{eq:mulambda}
    \mu_\Lambda = -\frac{K}{4\pi} \Lambda^2
\end{equation}
plus an unimportant constant, where $K$ is a classical parameter defined below. This is what we do always in this paper.
  \item[outer region $r>=r_0$.] In this region up to $\infty$ the solution of the radial equation is of the same form as Eq.~(\ref{eq:collisionradial}), but here the good quantum number is $N$ instead of $\Lambda$, so that it writes
\begin{equation}\label{eq:coulombradial}
    f_N(r) = \left( s(E_e ,r)\,c_N + c(E_e ,r)\,d_N \right)
\end{equation}
Furthermore, here $E_e$ is known as $E_e=E-B N (N+1)$, where $B$ is the rotational constant of the core, according to standard molecular conventions \cite{HerzbergI50}.
  \item[matching at $r=r_0$.] These two solutions are two developments of the same wavefunction on two different angular basis.
      Matching involves an overlap matrix
\begin{equation}\label{eq:Udef}
    X^{(L J)}_{\Lambda,M_J}=\sum_N\Phi^{(L J)}_{N,M_J}
\matr{U}_{N\Lambda}^{(LJ)},
\end{equation}
which is proportional to a Clebsch-Gordan coefficient \cite{fano70}:
\begin{equation}\label{eq:Uclebsch}
    \matr{U}_{N\Lambda }^{(LJ)}=\langle L\,-\Lambda, J\,\Lambda\vert L J N 0\rangle
(-1)^{J-N+\Lambda },
\end{equation}

Fano \cite{fano70} gives also an extra symmetrization on $\pm\Lambda$ due to the symmetry of the potential by reflection onto the $\vec{M}$ axis.

  \item[quantization.] It is obtained by demanding that the growing Coulomb solution at $r\to\infty$ be zero, which is obtained if energy $E$ is such that Eq.~(\ref{eq:coulombradial}) reduces to \cite{seaton83}
\begin{equation}\label{eq:radialinfty}
    f_N(r) \propto -\cos(\pi\nu_N)\,s(E_e,r) + \sin(\pi\nu_N)\,c(E_e,r),
\end{equation}
where $\nu_N$ is the principal quantum number of the electron in channel $N$ given in atomic units by

\begin{equation}\label{eq:nuN}
    E=B N (N+1)+E_e =B N (N+1)-\frac{1}{2\nu_N^2}
\end{equation}
\end{description}

The final quantization set of equations is
\begin{equation}\label{eq:EqClosed}
\begin{array}{lcr}
  \sum_\Lambda \matr{U}_{N\Lambda} \sin(\pi(\nu_N+\mu_\Lambda))A_\Lambda = 0
  &;& N=J-L \cdots J+L
\end{array}
\end{equation}
The standard way to solve it is to look to zeros of the preceding determinant (and thus non zero values of the coefficients $A_\Lambda$) while varying total energy $E$, and thus $\nu_N$ and possibly $\mu_\Lambda$ which depend on $E$.

\bibliography{paper}

\end{document}